\documentclass[aps,prx,reprint,groupaddress,floatfix,balancelastpage]{revtex4-1}
\usepackage[breaklinks=true, colorlinks=true, linkcolor=blue, urlcolor=blue, citecolor=blue]{hyperref}

\usepackage{amsfonts}
\usepackage{amsmath}
\usepackage{color}
\usepackage{graphicx}
\usepackage{placeins}
\usepackage{subfigure}
\usepackage{soul}
\usepackage{ulem}

\def\bra#1{\langle{#1}|}
\def\ket#1{|{#1}\rangle}

{\catcode`\|=\active
  \gdef\Braket#1{\begingroup
\mathcode`\|32768\let|\BraVert\left<{#1}\right>\endgroup}}
\def\BraVert{\egroup\,\mid\,\bgroup}

\begin{document}

\title{Supersensitive measurement using single-atom control of an atomic ensemble}

\author{Calum MacCormick}
\affiliation{The Open University, Walton Hall, MK7 6AA, Milton Keynes, UK}

\author{Silvia Bergamini}
\affiliation{The Open University, Walton Hall, MK7 6AA, Milton Keynes, UK}

\author{Chris Mansell}
\affiliation{The Open University, Walton Hall, MK7 6AA, Milton Keynes, UK}

\author{Hugo Cable}
\affiliation{Centre for Quantum Photonics, H. H. Wills Physics Laboratory and Department of Electrical and Electronic Engineering, University of Bristol, Merchant Venturers Building, Woodland Road, Bristol BS8 1UB, UK}

\author{Kavan Modi}
\affiliation{School of Physics \& Astronomy, Monash University, Victoria 3800, Australia}

\date{\today}

\begin{abstract}
We analyze the operation of a novel sensor based on atom interferometry, which can achieve supra-classical sensitivity by exploiting quantum correlations in mixed states of many qubits. The interferometer is based on quantum gates which use coherently-controlled Rydberg interactions between a single atom (which acts as a control qubit) and an atomic ensemble (which provides register qubits). In principle, our scheme can achieve precision scaling with the size of the ensemble --- which can extend to large numbers of atoms --- while using only single-qubit operations on the control and bulk operations on the ensemble. We investigate realistic implementation of the interferometer, and our main aim is to develop a new approach to quantum metrology that can achieve quantum-enhanced measurement precision by exploiting coherent operations on large impure quantum states.  We propose an experiment to demonstrate the enhanced sensitivity of the protocol, and to investigate a transition from classical to supra-classical sensitivity which occurs when using highly-mixed probe states.
\end{abstract}


\maketitle


A key new quantum technology is quantum metrology, which encompasses techniques for measuring unknown physical parameters at the limits imposed by quantum mechanics \cite{Giovannetti04}. When physical parameters are estimated using traditional interferometric techniques based on single-particle probe states, the measurement precision is subject to the standard quantum limit (SQL) (also called the shot-noise limit) for which precision scales as $1/\sqrt{\nu}$, where $\nu$ is the number of repeated measurements. A large amount of research has focused on achieving precision beyond the SQL using non-classical probe states \cite{GLM11}. In particular, it is well known that path-entangled NOON states, which are states of the form $\left( \ket{N0} + \ket{0N} \right)/\sqrt{2}$ in the Fock basis \cite{DowlingReview}, can be used to attain the ultimate limit to measurement precision, which is called the Heisenberg limit, for which the precision after $\nu$ repetitions scales as $1/\sqrt{\nu} N$.  Equivalently, $N$-qubit Greenberger-Horne-Zeilinger (GHZ) probe states can be used to achieve the Heisenberg limit in a quantum circuit in analogy to an interferometric setup \cite{CavesShaji10}. However, attempts to achieve the Heisenberg limit using various experimental platforms are hindered by the difficulties of generating large NOON or GHZ states with near-perfect purity. For example, the largest NOON states reported so far in optical interferometry comprised five photons \cite{Afek2010}. Recent theoretical work has also shown that particle losses make quadratic enhancement of precision impossible to achieve as the probe-state particle number increases to large values \cite{RafalNatComms12}. Larger NOON states have been demonstrated for ensemble-spin systems in nuclear magnetic resonance (NMR) experiments, although the scalability of these systems is limited by molecules with fixed numbers of spins \cite{Jones2009}.

A new perspective on the resources required to achieve quantum advantage for measurement precision is provided by two theoretical works which investigate mixed-state models of quantum metrology \cite{arXiv:1003.1174, hugo}.  Ref.~\cite{arXiv:1003.1174} compares quantum probes which can be prepared using unitary circuits from initial qubits which are mixed, uncorrelated and identical. For a given number of qubits and fixed purity, it was found that circuits generating quantum correlations can achieve a quadratic enhancement in measurement precision (compared to the use of the initial qubits as separate probes). This result remains true even when the qubits are highly mixed and the relevant circuit generates no entanglement, which demonstrates that pure-state entanglement is not a pre-requisite for quantum enhancements in metrology. Considerable evidence was also found for an important role for quantum discord \cite{Modi12}, which does remain when the qubits are highly mixed. More recently, a model of quantum metrology inspired by the one-qubit quantum-computation model DQC1 \cite{KnillLaflamme98} was analyzed in Ref.~\cite{hugo}. This model uses input states comprising one pure qubit together with a register of fully-mixed qubits.  Surprisingly, the SQL is achieved in this model even though the amount of coherence is vanishing when the register is arbitrarily large. The SQL is then exceeded using an additional qubit which provides only a small amount of additional purity, and which leads to some nonclassical correlations between probe qubits.

In this paper, we study theoretically a realistic implementation of a new scheme for high-precision measurement (beyond the SQL) using a cold-atoms architecture which can practically access a large resource of readily-available mixed qubits.  The scheme is related to those in Refs.~\cite{arXiv:1003.1174, hugo}, and the specific implementation we consider is based on quantum correlations between a single ``clean'' (nearly-pure) control qubit and a register of partially-mixed register qubits, using a platform similar to the one described in Ref.~\cite{mansell}.  Correlations between the control and the mixed register are generated via a conditional quantum gate using a scheme first presented in Ref.~\cite{Muller2009}.  Our cold-atom architecture allows for ultra-precise measurement of important physical quantities including the strength of the local gravitational or magnetic field, or an applied oscillating electric field which irradiates the atoms, and crucially it can scale to a large number of register qubits while maintaining coherent operation.  In particular, the number of register qubits can be substantially larger than that achieved in experiments based on NMR, but with a purity approaching that of photonic or ion-based systems.

We proceed as follows: In Sec.~\ref{sec:Protocol} we explain the protocol and the achievable sensitivity.  Next in Sec.~\ref{sec:Implementation} we provide a detailed analysis of implementation of the protocol, taking into account all major experimental imperfections which are Poissonian fluctuations of the number of register atoms, atom losses from dipole traps, spatial distributions of the trapped atoms, and intra-register interactions.  Finally, in Sec.~\ref{sec:Discussion}, we propose an application of our scheme to measuring gravity using state-dependent gravitational potentials.

\begin{figure*}[t]
\includegraphics[width=0.9\textwidth]{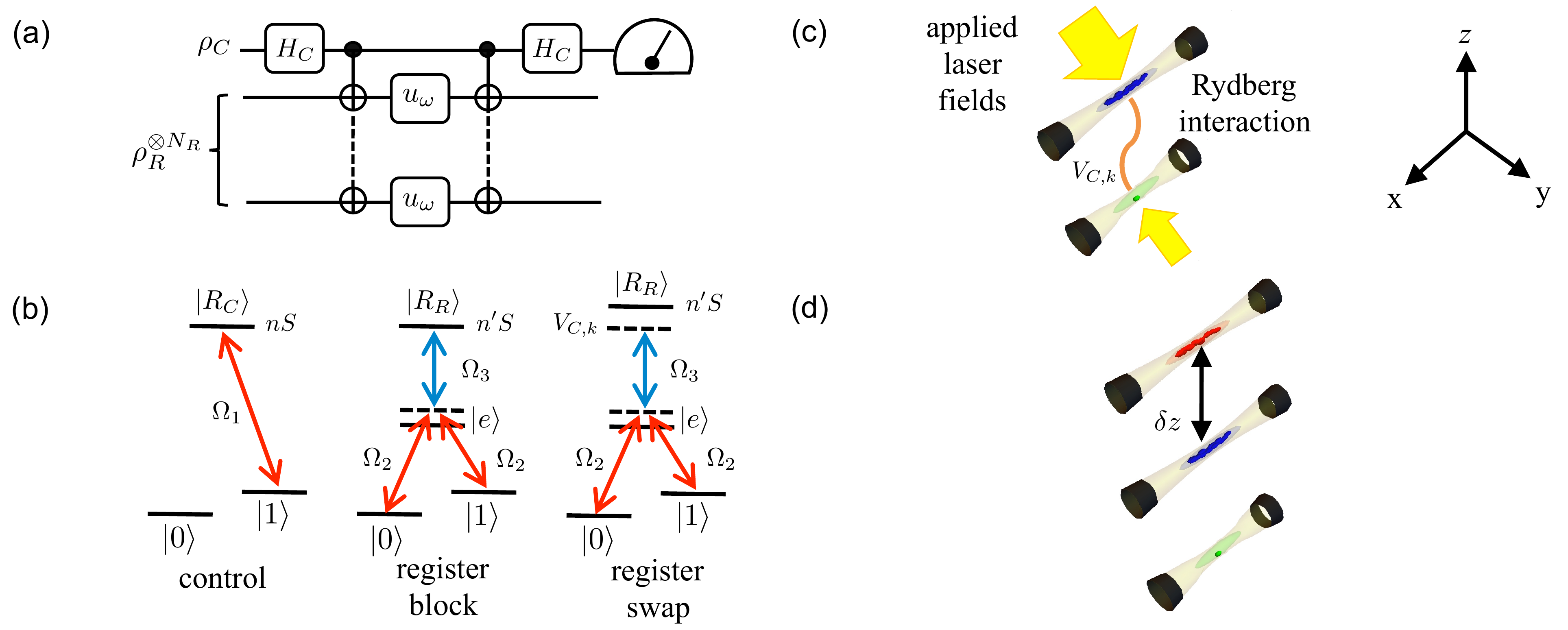}
\caption{(a) {\it The protocol:} The control qubit is prepared with high purity in state $\ket{0}$ followed by a Hadamard gate ($H_C$), and a $\textsc{Cnot}$ gate prepares the probe state.  Next the interaction $u_\omega$ is applied, driving the evolution of the probe state. The $\textsc{Cnot}$ is reapplied and finally the control atom's state is measured. (b) {\it Level schemes:} The four-level schemes for the control and register atoms. The control atom is coupled to the Rydberg state $\ket{R_C}$ via a laser pulse. The register atoms must be coupled by three lasers, far-detuned from the intermediate $\ket{e}$, but Raman resonant with the (unperturbed) Rydberg levels. The $\ket{0}\rightarrow\ket{e}$ and $\ket{1}\rightarrow\ket{e}$ drive a $\pi$ pulse between the $\ket{0}$ and $\ket{1}$ states only when the electromagnetically-induced transparency (EIT) condition is lifted by the strong Rydberg-Rydberg interaction due to the control atom in state $\ket{R_C}$. (c) {\it Implementation:} In the experiment, the tight control trap will contain only a single atom while the larger register trap will be populated with an average of 25 atoms.  The atomic positions sample the density distribution for the traps as described in the text. (d) {\it Gravity measurement:} Implementation of a gravity measurement may be accomplished by using independently-controlled and state-dependent optical potentials which allow the register trap to be divided along the $z$ axis ($\bf{g}$ lies along $z$). \label{fig:scheme}}
\end{figure*}

\section{The protocol}\label{sec:Protocol}

In this section, we describe our protocol in a generic fashion, following Fig.~\ref{fig:scheme}(a). We assume a single ``clean'' qubit called the control, upon which we can make high-fidelity single-qubit operations and high-efficiency measurements.  In addition, we assume a register of $N_R$ qubits which are not individually addressable, but upon which it is possible to make bulk operations.  The control and register qubits are initialized in the state
\begin{gather}
\label{eqn:1x}
\rho_{C,R} = \left( \frac{1+p_{C,R}}{2}\right) \ket{0}\bra{0} +\left(
\frac{1-p_{C,R}}{2}\right) \ket{1}\bra{1},
\end{gather}
where the subscript $C$ applies to the control qubit and $R$ to the register qubits, and where $0 \leq p_{C,R} \leq 1$ corresponds to noise in each control/register qubit. After initialisation, a Hadamard gate puts the control qubit into the state:
\begin{gather}
\label{eqn:controlHad}
\rho_C = \frac{1}{2}\left(\ket{0}\bra{0} +\ket{1}\bra{1}\right) + \frac{p_C}{2}\left(\ket{0}\bra{1}+\ket{1}\bra{0}\right).
\end{gather}
The preparation of the ``probe'' state is completed by a $\textsc{Cnot}$ which flips the register qubits conditional on the control qubit. Following this gate, the global state of the probe is
\begin{align}
\label{eqn:probestate}
\rho  =& \sum_{{\bf x}_n}
\frac{\left( 1\;+\;p_R\right) ^{N_{R}-n}\;\left( 1\;-\;p_R\right)^n}{2^{N_{R}}} \nonumber\\
& \quad \times \left(\frac{1+p_C}{2} \ket{g^+_n}\bra{g^+_n}
+\frac{1-p_C}{2} \ket{g^-_n}\bra{g^-_n}\right).
\end{align}
Above $\ket{g_n^\pm} = ({\ket{0 \, {\bf x}_n} \; \pm \;\ket{1\, {\bf x}_n^{c}}})/\sqrt{2}$, where $n$ is the number of $1$s in ${\bf x}_n$, complementary string ${\bf x}_n^{c}$ is the $\textsc{not}$ of ${\bf x}_n$, and the sum is taken over bit strings ${\bf x}_n$ of length $N_{R}$.

Next, the probe evolves under interaction with an unknown field for a time $t$. This interaction is assumed to be unitary, and to act on each qubit independently, i.e., $u_\omega^{\otimes N_R}$. The unitary transformation has the form $u_\omega = \exp(-i \, \omega H t)$, with controllable interaction time $t$, Hamiltonian $H = \ket{1}\bra{1}$, and unknown coupling parameter $\omega$ (which we wish to estimate). This interaction causes evolution of the probe state within orthogonal two-dimensional subspaces, for which
\begin{align}
u_\omega^{\otimes N_R} \ket{g^\pm_n} =& \cos\left[\left(\frac{N_R}{2} -n \right) \omega t \right] \ket{g^\pm_n} \\\notag
&+ i \sin\left[\left(\frac{N_R}{2} -n \right) \omega t \right] \ket{g^\mp_n}.
\end{align}
To extract information about $\omega$ the first two steps are applied in reverse, preparing the system for the measurement stage. The measurement outcomes have probabilities
\begin{align}
\label{eq:FringeProbability}
P_{k,n} =& \binom{N_{R}}{n} \frac{\left( 1+p_R\right) ^{N_{R}-n}\left( 1-p_R\right)^{n}}{2^{N_{R}}}  \nonumber\\
& \times \frac{1+ (-1)^k p_C }{2}\, \cos\left[(N_R - 2n) \omega t\right],
\end{align}
where $P_{k,n}$ is the probability of measuring $n$ register qubits in state $\ket{1}$ along with the control qubit in state $\ket{k},\,k \in \{0,1\}$.

Information about the sensitivity for estimating $\omega$ can be extracted from the probability distributions above. Formally this can be done by computing the Fisher information, $F(\omega) =\sum_{k,n} \left(\partial_\omega {P_{k,n}} \right)^{2} /{P_{k,n}}$, which is defined for arbitrary probability distributions $P_{k,n}$.  $F\!\left( \omega \right)$ gives the precision for estimating the unknown $\omega$ using an efficient and unbiased statistical estimator as $\Delta \omega = 1/\sqrt{\nu F}$ (corresponding to statistical estimation that saturates the Cram\'er–--Rao bound) where $\nu $ is the number of repeated measurements.  In this paper, we define {\it sensitivity} as
\begin{gather}\label{sens}
\frac{\delta\omega}{\omega}=\frac{1}{\omega \sqrt{\nu F(\omega)}}.
\end{gather}

For our protocol, the Fisher information for the case $p_C=1$ is given by
\begin{gather}
\label{eq:IdealisedFisherInfo}
F=\left[ \left( 1-p_R^{2}\right) N_{R}+p_R^{2}N_{R}^{2}\right] t^{2},
\end{gather}
and there is no dependence on $\omega$.  It can also be verified that the protocol here is optimal in the sense of Ref.~\cite{CavesBraunstein94}. When the register is fully pure ($p_R=1$), we recover the Heisenberg limit for precision
$\Delta \omega \ge 1/\sqrt{\nu }N_{R}t$ (recall that the control itself does not interact with the field). When instead the register qubits are fully mixed at the input ($p_R=0$) we find that $\Delta \omega \ge 1/\sqrt{\nu N_{R}} t$, which is the SQL. Hence any finite purity yields supra-classical sensitivity \cite{hugo}, and a large qubit ensemble is seen to become a powerful resource for parameter estimation when supplemented with the coherence originating from one ``clean'' qubit.

From a practical perspective, measurement of all the qubits in a large register is in general challenging, and considerable simplification of our proposal is achieved by restricting the measurement to the state of the control qubit only. On the other hand, the control qubit itself  might suffer a small loss of purity so that $p_C<1$. The measurement probabilities for the control qubit are now given by
\begin{gather} \label{eq:P0P1}
P_{0} = \sum_n P_{0,n} \quad \mbox{and} \quad
P_{1} = \sum_n P_{1,n}.
\end{gather}
Fisher information derived from $P_0$ an $P_1$ is in general lower than that of Eq.~\eqref{eq:IdealisedFisherInfo} and depends on the value of the unknown $\omega$. An adaptive process is therefore required to implement offsets $\omega \mapsto \omega +\omega _{\rm ad}$ over a series of measurements, to tune the circuit to the region of greatest sensitivity. Returning to Eq.~\eqref{eq:P0P1}, the effect of $p_C < 1$ is seen to reduce fringe visibility. For the special case that $p_C=1$, $F(\omega)$ is maximized at $\omega t=\pi a$ (where $a$ is integer) where it attains the Fisher information Eq.~\eqref{eq:IdealisedFisherInfo}. This however relies on the possibility for perfect extinction of $P_{0}$ or $P_{1}$, and
corresponds to stationary points for $P_{0}$ and $P_{1}$.  In general $p_C <1$, and the device is not sensitive to the precise value of $\omega $ at these points ($F=0$); instead $F$ is maximized at points close to $\omega t=\pi a$ where the gradients of $P_{0}$ and $P_{1}$ are steepest.

Finally, just like NOON and GHZ states, the probe states created in our protocol, Eq.~\eqref{eqn:probestate}, are extremely fragile with respect to loss: the removal of a single qubit from the probe state, before or while it interacts with the unknown external field, destroys all coherence in the GHZ-state components. Consequently, loss of a single qubit causes $P_0$ and $P_1$ not to depend on the value of $\omega$, and all useful information is destroyed.

\section{Implementation}
\label{sec:Implementation}

\subsection{Experimental setup}
\label{sec:ExperimentalSetup}

In this section we discuss implementation of our protocol from Sec.~\ref{sec:Protocol} using a cold-atoms-based platform.
As shown in Fig.~\ref{fig:scheme}(c), in our proposed implementation, $^{87}$Rb atoms are trapped in two individually addressable and tightly-confining optical-dipole traps, one containing a single control atom and the other the register atoms \cite{mansell, Muller2009}.  These can be realised from a single laser beam using spatial light modulators \cite{bergamini} or acousto-optic deflectors. These sort of traps offer many advantages for the scheme including readily-reconfigurable potentials  as well as micron-scale separation between control and register qubits.

The implementation of the $\textsc{Cnot}$  gates to prepare the probe state is based upon the proposal in Ref.~\cite{Muller2009}, which uses electromagnetically-induced transparency (EIT) and Rydberg blockade. Controlled gates based around the strong interactions between Rydberg atoms are ideal for quantum metrology, as the strong and laser-tuned long-ranged interactions allow for great control contrast and high gating fidelity.

Note that our approach here to building conditional gates has some drawbacks: when working with large register sizes, high atomic density may lead to large collisional phase shifts which demand additional experimental complexity to mitigate (for example, suppressing the collisional interactions by tuning a magnetic field close to a Feshbach resonance). In optical traps, the density should be limited to around $10^{12} \;\mathrm{cm}^{-3}$ to avoid excessive trap losses due to light-induced collisions. We include discussion of the limitations imposed by density-dependent losses below.

Trapping potentials suitable for this scheme are provided by a pair of far-detuned optical trap (FORTs) at 825 nm, propagating along the $x$-axis, as shown in Fig.~\ref{fig:scheme}(c,d) and focused to a waist of $w_0=1.0 \; \mu\mathrm{m}$ for the control atom. The chosen trap depth is 1 mK: with a temperature of $100 \; \mu \mathrm{K}$ the atom is confined to a volume with radial and axial widths $0.08 \; \mu\mathrm{m}$ and $0.30 \; \mu\mathrm{m}$. For the register, it is necessary to restrict the density of the atoms without allowing the atoms to travel much further than the $r_\mathrm{max} = 6\; \mathrm{\mu m}$ range of the Rydberg-Rydberg interaction (see below).  We can achieve a reasonable trap using an elliptical beam profile with waists of $w_{z,0} = 1.2\; \mathrm{\mu m}$ and $w_{y,0}=10\; \mathrm{\mu m}$, where $w_{z,0}$ is the beam waist measured along the gravitational axis (or $z$-axis in figure), and $w_{y,0}$ is the waist along the axis $y$, as in Fig.~\ref{fig:scheme}. The traps are separated by $2\; \mathrm{\mu m}$. The ensemble atoms also have a temperature of $100 \; \mu \mathrm{K}$ so the atoms are confined in an oblate-ellipsoid shape with widths of $s_x = 1.73 \;\mathrm{\mu m}$, $s_y = 1.58\;\mathrm{\mu m}$ and $s_z = 0.19 \;\mathrm{\mu m}$ along the $x$, $y$ and $z$-axis respectively, for a density of $1.24 \times N_R \; \times 10^{11}\mathrm{cm^{-3}}$.

The qubits are encoded in the $5S_{1/2}$ ground state hyperfine sublevels with $F=1$ being assigned to $\ket{0}$ and $F=2$ assigned to $\ket{1}$. To implement the gates, the traps are illuminated by focused laser beams, which couple the four relevant states of the atoms as shown in Fig.~\ref{fig:scheme}(b). It is of crucial importance that the beams which generate rotations are of uniform intensity for all target  atoms; this can be achieved by ``flat-top'' profile beam-shaping techniques \cite{ReetzLamour08}. To initialise the atoms we will follow the protocol given in Ref.~\cite{mansell}, where a Rydberg-atom-based DQC1 implementation was proposed. In contrast to the DQC1 protocol of Ref.~\cite{mansell}, in which a requirement is preparation of each register qubit in the mixed state $\openone/2$, here we need only to prepare the atoms in the state $\ket{0}$ with as high purity as is attainable, and this can be performed using optical pumping.

\subsection{Noise sources and practical limitations}
\label{sec:NoiseSources}

We now discuss the practical limitations which affect the implementation of our protocol and our strategies to mitigate the effects. The trap lifetime limits the duration over which the evolution in the unknown field can extend, whilst atom-number fluctuations for the register-trap loading affect the overall contrast and shape of the fringes.  Another important consideration is the achievable fidelity of the gate operations, which are limited by laser performance (linewidth) and the uniformity of the intensity profile over the register ensemble.

{\it Atom losses:} The most severe constraint on the protocol is imposed by atom losses. As noted above in Sec.~\ref{sec:Protocol}, the probe state is extremely fragile with respect to atom loss or environmental disturbance: the removal of a single qubit from the register in a run of the experiment destroys all information about the parameter $\omega$, so that measurement on the control qubit yields either of two outcomes with equal probability. Therefore, averaging over $\nu$ runs,  atom losses lead to reduced fringe contrast, in line with the probability that a single atom is lost from the experiment.  The improvement in sensitivity with register size is limited by the commensurate increasing probability to lose an atom during the time between the controlled gate operations, and there are therefore strong constraints on both the register size and the experimental timescales.

\begin{figure*}[t!h!]
\includegraphics[scale=0.44]{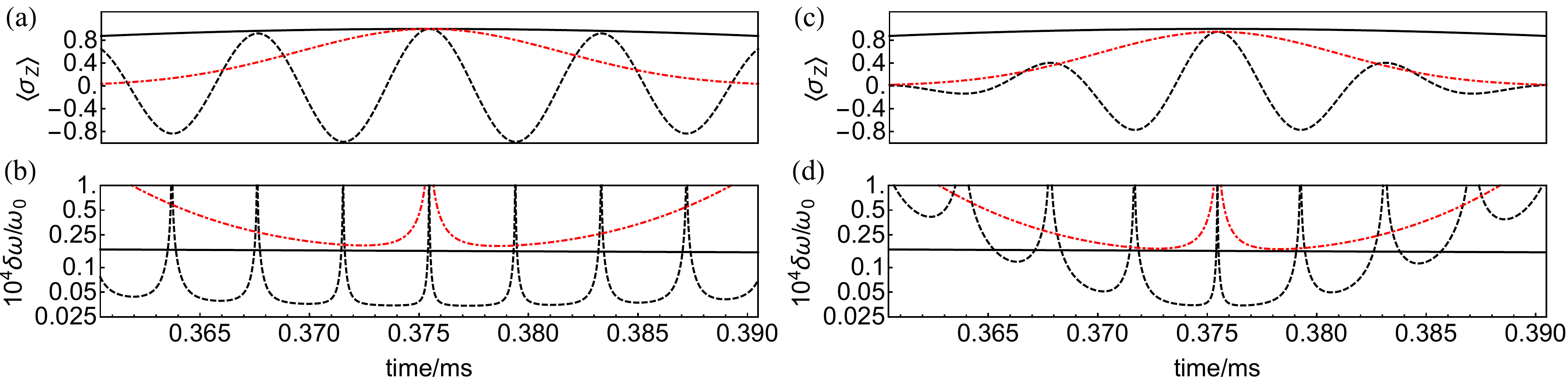}
\caption{{\it Dephasing of the interference fringes due to reduced register purity and register-size fluctuations:} (a) Interference fringes in $\sigma_Z$ as $t$ is scanned with $N_R=25$, $p_C=1$; $p_R=0.95$ (black dashed line) and $p_R=0$ (red dot-dashed line); the black solid line shows the fringe for each of 25 perfectly-pure qubits (for comparison to a perfect classical interferometer); (b) The sensitivity obtained via the Fisher information for the fringes obtained in (a) using the same colors; (c)  Interference fringes corresponding to the inclusion of Poissonian fluctuations in $N_R$, which lead to further dephasing of the fringes; (d) The sensitivity obtained from the fringes in (c). \label{fig:ReducedPurityFluctuations}}
\end{figure*}

Our proposed implementation assumes that optical dipole traps are used for the atoms.  Consequently, limitations are caused by trap losses due to imperfect vacuum (which can allow for trap lifetimes of $\approx$ 1 minute at pressures $\approx 1\times 10^{-11}\;\mathrm{mBar}$), and also by spontaneous Raman transitions which effectively make an unwanted measurement of the state of the system.  In addition, a density-dependent two-atom loss rate for an optical dipole trap is given by $\gamma_2 = -\beta \int n(r)^2 dV$, where $n(r)$ is the atomic density as a  function of position $r$, and the value of the constant $\beta$ strongly depends on the trap laser. For the gaussian density distribution of the register,
\begin{gather}
\gamma_2 = -\frac{\beta N_R(N_R-1)}{8\pi^{3/2} s_x s_y s_z}.
\end{gather}
The timescale for trap-laser-induced Raman transitions has been reported as being as long as $\tau_{sp}=3.3\;\mathrm{s}$ \cite{Frese2000}, and in the same experiments $-\beta \int n(r)^2  < 0.02\; \mathrm{s^{-1}} $, from which we obtain an estimate of $\beta=0.25 \times 10^{-12} \;\mathrm{cm^{3} \;s^{-1}}$ and we may estimate $\gamma_2 = 6.47 \; s^{-1} $ for the ensemble trap with $N_R=25$ atoms. Using these quantities, we estimate that the probability to not lose a single register atom is dominated by the two-body loss, so that $P_\mathrm{no-loss}=\exp(-N_R t \gamma_2)$.  For this reason, we have chosen a register size of $N_R=25$ atoms, which, when compared to the classical interferometry, allows a significant improvement in the sensitivity with a probe evolution time of $375\;\mathrm{\mu s}$ and $P_\mathrm{no-loss}\approx 0.94 $.

{\it Atom-number fluctuations:} The register size will be subject to number fluctuations due to the trap-loading process, which leads to a dephasing of experimental fringe patterns. As an example, consider the simplified case with $p_C\!=\!1$ and $p_R\!=\!1$, where the number of atoms in the ensemble are sampled from the Poisson distribution with mean $N_R$: in this case the fringes recovered at each run will sample from
\begin{gather}
\frac{1}{2}\pm \frac{e^{-N_R}}{2} \sum_{m=0}^\infty \frac{N_R^m}{m!} \cos(m \omega t).
\end{gather}
The fringes and the corresponding Fisher information must be obtained from the probabilities $P_{k,n}$ which depend upon $p_C$, $p_R$ and $N_R$.

The effects of register-size fluctuations are illustrated by Fig.~\ref{fig:ReducedPurityFluctuations}, which shows interference fringes and corresponding sensitivity for the case when $\omega = 2\pi\times 5.33\;\mathrm{kHz}$ and $t\approx 375\;\mathrm{\mu s}$ with reduced register purity, before and after the fluctuations are accounted for (Fig.~\ref{fig:ReducedPurityFluctuations}(a,b) and Fig.~\ref{fig:ReducedPurityFluctuations}(c,d) respectively). Although the envelope of the fringes is narrowed by the fluctuations in register size, supra-classical precision is possible: for a mean register size of $N_R=25$, sensitivity (extracted form the Fisher information) is achieved equal to that of a SQL device (operating with 25 \textit{pure} single qubits) when the purity of the register is as small as $p_R = 0.09$.  We conclude that the performance would not be seriously hampered by fluctuations in register size as long as the interferometer is operated when $\omega t$ is close to an integer multiple of $2 \pi$.

\textit{Intra-register interaction and gate errors:}
We now proceed to numerical studies of the performance of our protocol.  Detailed modelling is needed to determine gate fidelity, and to find optimal values for experimental parameters.  Unfortunately, modeling of the full protocol requires solving a master equation for $N_R+1$ four-level atoms, and this is impractical for $N_R=25$.  Instead, we study first a complete simulation for systems of up to three register atoms ignoring the Poisson loading statistics, a simplified case for which we can solve the equations of motion, before discussing the $N_R=25$ case using simplified modelling.  Although experiments involving three register atoms would not make for a quantum-enhanced sensor, they would be sufficient to analyze the effects of interactions between ensemble atoms, and can be rapidly implemented and rigorously understood in state-of-the-art experiments (unlike larger systems).  Of particular importance is understanding the effects of varying Rydberg-Rydberg interaction strength due to the atomic density distribution, which potentially modulates the gate performance especially for register atoms located far from the control atom, as well as confirming whether the effects of the intra-register interactions are negligible in the gating operation.

We evaluated the performance of the Rydberg-based interferometer by numerically simulating the whole protocol comprised of two $\textsc{Cnot}$ gates and a given applied phase shift, which is the parameter we want to estimate.  The details for our simulation are as follows (further details are also given in the Appendix \ref{app:sol}):  As mentioned earlier, the gate scheme is based on the proposal in Ref.~\cite{Muller2009}, with the atomic level scheme and wavelengths involved represented in Fig.~\ref{fig:scheme}(b).  We consider each atom as a four-level system, consisting of the  hyperfine levels of the ground state with  $5S_{1/2}$, F=1 being $\ket{0}$ and F=2 $\ket{1}$, the excited $5P_{3/2}$ state $\ket{e}$ and a Rydberg state $\ket{R_{C,R}}$.  As in Ref.~\cite{Muller2009}, we choose to shape the laser pulses  such that
\begin{gather}
\Omega_2(t)=\sqrt{\frac{8\Delta_e}{3\tau}}\sin^2 \left( \frac{\pi t}{\tau} \right) \quad \mbox{and} \quad \Omega_3 = 10\sqrt{\frac{4\Delta_e}{3\tau}}
\end{gather}
with $\Delta_e = 2\pi \times 1000 \;\mathrm{MHz}$ and $\tau=0.5\;\mathrm{\mu s}$, which guarantee that the pulse area of the Raman transition is $\pi$, that the gate time is much shorter than the lifetime of the Rydberg states ($\approx 1\mathrm{ms}$) and that high fidelity gate performance is obtained.

Efficient Rydberg blockading is achieved when the Rydberg-Rydberg interaction between control and ensemble is much stronger than the Rabi frequencies involved in the gating step. When this condition is fulfilled, the fidelity of the gate does not depend on small perturbations of the interaction strengths \cite{Muller2009}. We therefore adopt a simplified treatment of the Rydberg-Rydberg interactions, so that we only consider the strongest interaction channel between two atoms excited to the $n \, S \, n^{\prime} \, S$ state, as discussed in \cite{saff10}. The levels considered and the definitions of parameters are discussed in Appendix \ref{app:spa}.

When the atoms are sufficiently close, the interaction scales as $r^{-3}$, turning over to an $r^{-6}$ beyond a certain range $r_{\mathrm{max}}^3=2C_{\mathrm{dd}}/\Delta_{\mathrm{def}}$, where $C_{\mathrm{dd}}$ is the reduced matrix element as defined in Eq.~\eqref{eq:cdd} and $\Delta_{\mathrm{def}}$ is the energy defect between two coupled channels. Exciting the control atoms with $n=74$ and the ensemble atoms with $n^{\prime}=73$, we have $C_{\mathrm{dd}}=2.92\times 10^4 \;\mathrm{MHz}\; \mathrm{\mu m}^3$ and $\Delta_{\mathrm{def}}=-196 \;\mathrm{MHz}$ for the control-ensemble interaction, yielding a range $r_{\mathrm{max}}=8.1 \;\mathrm{\mu m}$, and $C_{\mathrm{dd}}=2.84\times 10^4 \;\mathrm{MHz} \; \mathrm{\mu m}^3$, $\Delta_{\mathrm{def}}=-613 \;\mathrm{MHz}$ for the ensemble-ensemble interactions; the interactions are $r^{-3}$ in character for a distance of about $5.7 \mathrm{\mu m}$.

\begin{figure}[t]
\centering
\includegraphics[scale=0.35]{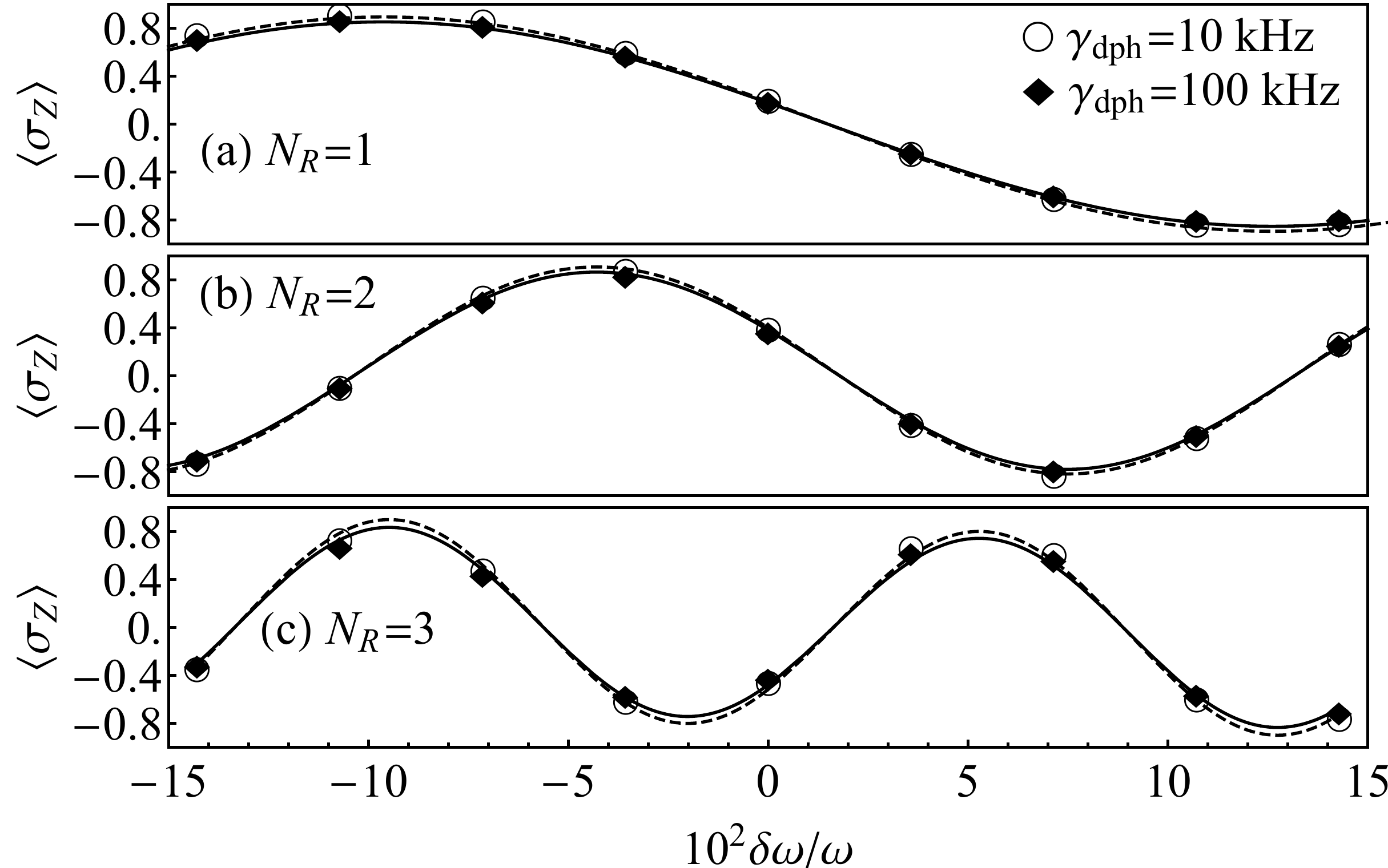}
\caption{{\it Results of the full density-matrix model of the protocol with $N_R=$1,2 or 3 atoms:} Interferometer ``output'' with finite laser linewidth, for $N_R=1$ atoms (a), $N_R=2$ atoms (b) and $N_R=3$ atoms (c). As described in the text, we have obtained each point by sampling the atomic density distribution for the position of control and ensemble atoms $\nu=49$ times, yielding $P_{0,1}(\omega)$ and hence $\langle\sigma_Z\rangle$ for each run. Due to the finite range of the Rydberg-Rydberg interaction, the gates have a range of fidelities which means there is a spread of $P_{0,1}(\omega)$ obtained at any given value of $\omega$. In each plot the purity of the control and register qubits are $p_C = p_R=0.95$ and we plot the average of the expectation values $\langle\sigma_Z\rangle$ obtained as a function of $\delta \omega/\omega_0$. The lines are fits to the output, from which we find that increasing the linewidth from 10 kHz to 100 kHz degrades the interferometer contrast from 0.928 to 0.908 in (a), from 0.896 to 0.886 in (b) and from 0.865 to 0.834 in (c).
\label{fig:Linewidth}}
\end{figure}

With the trap parameters as described in Sec.~\ref{sec:ExperimentalSetup}, we ran a Monte-Carlo simulation of the experiment. For each of the two applications of the gate, we picked atom positions at random from the appropriate atomic density distributions and then computed the strength of the interaction terms $V_{ij}(\bf{r}_{i,j})$ between atoms, $\bf{r}_{i,j}$ being the displacement vector between atoms $i$ and $j$. We then computed the time evolution of the system density matrix due to the full Hamiltonian. In between the gates, the atoms undergo evolution due to the applied field.  Finally, we applied a Hadamard gate to the control atom and traced over the register atoms, to yield the density matrix for the control atoms we wish to sample, thereby obtaining values for $P_0$ and $P_1$. We considered a number of repetitions $\nu=49$. For each of the two gates, we had 49 possible locations of each atom and obtained the expectation values of the operator $\sigma_Z$ corresponding to measuring the state of the control atom. The measurement was implemented by drawing, at random, the final measurement outcome from the binomial distribution values $P_0$ and $P_1$.  We also took into account the effect of laser linewidth, which reduces the fidelity of the gate.

The result of our simulations are shown in Fig.~\ref{fig:Linewidth}. We found that (for our choices of trap sizes and Rydberg levels) the performance of the protocol is dominated by the statistical uncertainty due to sampling the fringes $P(\omega)$  only $\nu$ times.  We therefore find that for an appropriate choice of the Rabi frequencies, the effects of the register-register interactions are negligible, and that the most significant cause of error in the gate is due to the small value of the Rydberg-Rydberg interaction strength when the distance between the control and register atom is large, such that $V_{C,k}\gg\hbar\Omega^2_3/4\Delta$ fails. That the register-register interactions have small effects is a surprising result described fully in Ref.~\cite{Muller2009}, and discussed more fully below in the context of systems where $N_R\approx 25$.

\begin{figure*}[ht]
\includegraphics[scale=0.4]{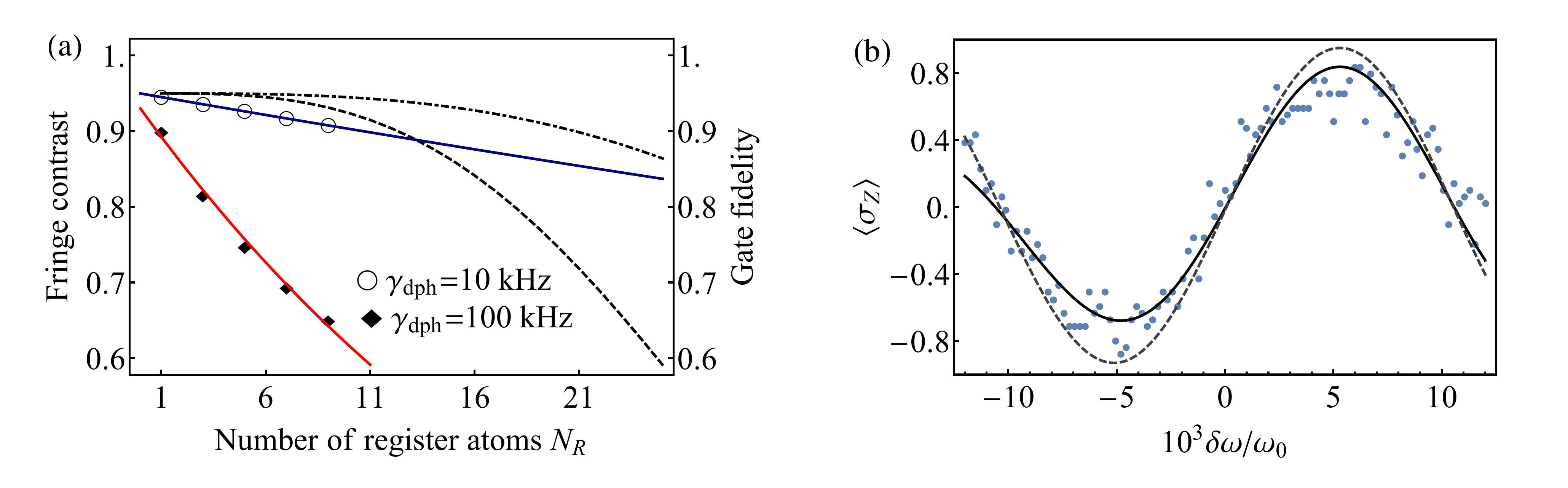}
\caption{{\it (a) Simulated effects of gate imperfections due to strong register-register interactions or weak control-register interactions:} The open circles show the falling gate {\it contrast} as more register atoms are added with 10 kHz laser linewidth; the filled diamonds for 100 kHz laser linewidth. An exponential fit allows us to extrapolate to a worst case scenario where for $N_R=25$ atoms the contrast has fallen to 0.75. The register-register interactions degrade the {\it fidelity} of the gate. The black dashed line shows the fidelity falling when $\Omega_3/\Omega_2=10$, and the black dot-dashed line when $\Omega_3/\Omega_2=15$, as calculated using the analytic form given in \cite{Muller2009}. {\it (b) Interferometer output with $N_R=25$ atoms when $\omega$ is scanned:} the purity of the control qubit and register qubits is 0.95, $t=375\;\mathrm{\mu s}$, and we scan the frequency over the range $\pm\delta\omega = 0.15\time 10^{-3} \omega_0$, where $\omega_0 = 2\pi\times 5326 \;\mathrm{Hz}$. The points show the results of $\nu=49$ repetitions at each value of $\delta\omega/\omega_0$ obtained by including the falling gate fidelity in (a) and including the atom losses due to light-induced collisions. The solid black line shows the fringes in the limit $\nu\longrightarrow\infty$, but not taking into account atom losses, and the dashed line shows the fringes produced with perfect gate fidelity.
\label{figure:includeFidelity}}
\end{figure*}

Regarding requirements on laser linewidth, typical laboratory set-ups can obtain laser linewidths of $\approx 100\;\mathrm{kHz}$ with standard laser locking techniques. For our proposal much lower linewidths will be required, as its effects become significant for a larger number of atoms in the probe. In Fig.~\ref{fig:Linewidth}, we show the interferometer output for the cases where the appropriate laser linewidth $\gamma_{dph}$ takes values $\gamma_{dph}=10.0 \;\mathrm{kHz}$ and $\gamma_{dph}=100 \;\mathrm{kHz}$. In fact, by evaluating the fringe contrast for varying $\gamma_{dph}$, we find that the contrast decays exponentially according to $p_C \, \exp(-0.250 \;\mathrm{kHz}/\gamma_{dph})$, which implies that for fringes with contrast greater than $90\%$, laser linewidth should be limited to below 10 kHz (which should be considered very demanding).

We conclude this section with a discussion of the gate operation as more register atoms are added to the system, addressing decreasing gate fidelity as $N_R$ increases. Previously we have argued that the register-register atom interactions are (or can be controlled such that they are) unimportant, and we now justify this in two ways: Firstly, M\"uller et al. \cite{Muller2009} show that the register-register interaction degrades the gate fidelity due to the laser coupling of `gray' states which evolve a phase as the gate proceeds. In Fig.~\ref{figure:includeFidelity} (a) we show the effects on the fidelity in the worst case scenario, when all atoms experience $V_{j,k}\gg\hbar\Omega^2_3/4\Delta$. We find that as long as $\Omega_3/\Omega_2 >15$ the gate fidelity can be controlled: one must raise the coupling laser Rabi frequency $\Omega_3$ as high as possible. There are limits: apart from the technical considerations, the control-register interaction strength should obey $V_{C,k}\gg \hbar\Omega^2_3/4\Delta$. Secondly, we note that, when the above conditions are fulfilled,  inclusion of register-register interactions into our simulation actually always improves the gate fidelity. This is because the most significant degrading effect is due to the finiteness of the control-register interaction: in some instances, due to the spread in the spatial distribution of the atoms in the register trap, this interaction may become so weak that the EIT condition is not lifted and the $\textsc{Cnot}$ flip is inhibited. In such cases, the register-register interactions actually boost effectiveness by supplementing the control-register interaction \cite{Muller2009}.

We therefore model larger atom systems by assuming (a) that we may neglect the register-register interaction as per the discussion above, (b) that the imperfect $\textsc{Cnot}$ gate is accurately described by $\ket{0}\bra{0}\otimes I + \ket{1}\bra{1}\otimes U_k,$ where
\begin{gather}
U_k=a_k \sigma_X + b_k \sigma_Y + c_k\sigma_Z
\end{gather}
 and $|a|^2 + |b|^2+|c|^2 = 1$. The condition (b) follows from the neglect of the register-register interactions, in which case the EIT blocking effect works with high fidelity. We have checked that condition (b) is valid in our full simulations of 1-3 register atoms (as noted above, register-register interactions mostly improve the gate). In neglecting the interactions, we now have a simpler problem to solve, namely we need only compute the gate applied to register atom individually, and then construct the density matrix for the whole system. For each atom we can extract the $U_k$ and use this information to calculate the multi-atom gate. We apply a $\textsc{Cnot}$ gate to up to 9 atoms with positions picked at random from the atomic density distribution and then a second gate, with new positions, and setting the laser linewidth to 10 kHz. We repeat this $\nu=49$ times. In between the gates, the atoms evolve for a time $1/\omega_0$, such that systems with odd $N_R$ have $\langle \sigma_Z\rangle \approx -p_C$. This procedure captures the fringe contrast, and for perfect gates the consecutive $\textsc{Cnot}$ gates should yield a fringe contrast of $\approx p_C$. The results are summarized in Fig.~\ref{figure:includeFidelity}(a), where the contrast is plotted with increasing $N_R$. Taking an exponentially-decaying contrast as the worst case result of this analysis, we find that for $N_R=25$ atoms, the fringe contrast will drop to 0.84. Effectively, the gate imperfections considered here affect only the contrast of the fringes.

To further expand upon the model above, we summarize our findings as follows: we have found that the imperfect gates have $c\approx 0.1$ and $a-i b = 0.95 e^{-i\theta}$, so to a very good approximation, the gate imposes only an additional phase $\theta$ upon the phase evolution. Hence, in Eq.~\eqref{eq:FringeProbability}, the cosine terms are modified as follows: $ \cos[(N_R - 2n ) \omega t+ \sum_{g,j}{\theta^{(g)}_j} ]$, where $\theta^{(g)}_j$ is the extra phase pickup for the $j^{th}$ atom due to gate $g$, $g=1,2$, which tend to decrease the contrast of the fringes. As the cosines evolve more rapidly as $N_R$ increases, the measured fringes dephase more strongly with $N_R$.

Now we are in a position to model an actual experiment with a nominal $N_R=25$ atoms, and as noted above, the lifetime of the atoms in the trap will decrease with increasing $N_R$ and a single atom lost destroys all the phase information we wish to measure. For each of the $\nu=49$ instances of the simulated experiment, we pick $N_R$ from a Poisson distribution with a mean of 25 atoms, and then find the $\sigma_Z$ where we adjust the control purity according to Fig.~\ref{figure:includeFidelity}(a) and at random we lose an atom according to the probability ($N_R$ dependent) that a light-induced two-body loss occurs - thus taking $\sigma_Z=0$ for each case where an atom is lost. The results of this analysis are displayed in Fig.~\ref{figure:includeFidelity}(b) where we show how the fringes would look, accounting for atom losses and poor gate fidelity, but neglecting the register-register interactions. Obtaining the Fisher information from this experiment, we find that the sensitivity is $S_{\mathrm{Q}}=\delta\omega/\omega_0 = 4.24\times 10^{-3}$, whereas for a perfect classical device $S_{\mathrm{C}}=\delta\omega/\omega_0 = 15.8\times 10^{-3}$, the quantum-correlated device outperforming the classical device by a factor of $S_{\mathrm{C}}/S_{\mathrm{Q}} =3.73$.

\section{Discussion}\label{sec:Discussion}

{\it Gravity measurements:} Finally, we discuss the application of our protocol to measuring the local gravitational field. We assume that the register contains $N_R = 25$ atoms and that it follows Poissonian loading. The register atoms are loaded in a state-dependent trap: the ensemble trap can be engineered so that the atoms are guided by two state-dependent trapping potentials, which allow for atoms to shift vertically conditionally on their internal state \cite{Bloch, Meschede}. During the initialization and probe state preparation, the two potentials are overlapping. Subsequently, the two state-dependent traps are displaced along the $z-$axis (as illustrated in Fig.~\ref{fig:scheme}(c,d)) by an amount $\delta z$ relative to each other. We assume that we can adiabatically shift the two state-dependent potentials which confine the register atoms by $2.5 \;\mathrm{\mu m}$, as this can be efficiently done with no significant heating or atom losses, while preserving the qubit state with high fidelity, as shown in Ref.~\cite{Browaeys, Meschede, Bloch}. This enables splitting of the atom wave-function vertically by a distance $\delta z$, and a $\delta\phi = m g \delta z t/\hbar$ phase difference is induced by letting the probe evolve for a time $t$.

In the case of Rubidium, the value of $m$ above is $m_{Rb} = 1.45\times 10^{-25}\;\mathrm{Kg}$, and the phase parameter evolves at $\approx 2\pi\times 2145\;\mathrm{s^{-1}}$ per $\mathrm{\mu m}$ separation, thus amounting to $\omega = 2\pi\times 5326\; \mathrm{s^{-1}}$ for $\delta z=2.5 \; \mathrm{\mu m}$.  We thus consider the case where 25+1 atoms prepared in a probe state as in Eq.~\eqref{eqn:probestate}, with $p_C = 0.95$ and variable values for $p_R$.  Since the evolution time is strongly limited by trap losses due to light-induced collisions, the probe state is left to evolve under gravity for $375 \; \mathrm{\mu s}$, and hence the system is probed after two fringe periods. We have also assumed that the traps are shifted slow enough to retain the atoms without loss of coherence or increase in temperature. We may estimate this speed by noting that for a trap frequency of $f$ and a confinement length of $\sigma$, we need to move the traps no faster than $\delta z /\sigma f \approx300\; \mathrm{\mu s}$ as we move the atoms into place.

\textit{Conclusions:} In this paper, we have analyzed an interferometric protocol using cold atoms in which the purity of the probe state does not need to be close to 1, in order to gain sensitivity beyond the SQL which limits classical devices. We find that under reasonable experimental conditions $N_R=25$, $p_C=0.95$ and $p_R=0.95$, clear interference fringes can be obtained such that the performance (measured by the sensitivity) of a classical counterpart using $N_R$ independent qubits can be surpassed by a factor of $3.73$. The sensitivity of the protocol improves over the standard quantum limit with the number of correlated particles as $\sim\sqrt{N_R}$.  We have shown that the high sensitivity is obtained even though the fringe patterns are subject to dephasing due to finite purity and fluctuations in the number of register atoms, and we have derived the purity needed to surpass classical sensitivity in this cold-atoms-based experiment.

The experimental implementation we propose allows for a great range of tunability of the purities both of the control and ensemble atoms, as described in \cite{mansell}, making it possible to explore the regimes between ``classical'' and ``fully quantum'', and thereby gain an understanding of the role of quantum correlations in quantum-enhanced sensing.  Lastly, we make a comment about the situations in which such a super-sensitive device would be advantageous in practical use:  Although strongly limited by two-body losses, our scheme could be of utility in situations where rapid sampling of a field is required (cooling and trap-loading requirements are far simpler compared to BEC setups for example). The sensitivity of a classical interferometer with $N$ atoms and an interaction time $t$ is limited to $1/t\sqrt{N}$, which shows that lengthening $t$ is the cheapest way of improving the device. However, for situations where rapid sampling is required, $t$ will be compromised --- and in fact, so too will $N$ due to shortening of the time taken to load a large number of ultra-cold atoms in a MOT --- precisely the situation in which our proposed scheme operates.

\acknowledgements
CMC, SB, and CM would like to acknowledge the support of EPSRC grant number EP/K022938/1. HC would like to acknowledge funding from the University of Bristol.
\appendix

\section{Solving the master equation for up to three register atoms}\label{app:sol}

The single-atom Hamiltonian is
\begin{align}
H_k=&\Delta_e  \ket{e}_k \bra{e} +\frac{1}{2}\Omega_2 \left(\ket{0}_k \bra{e} + \ket{1}_k \bra{e} \right) \nonumber\\
&+ \frac{1}{2}\Omega_3 \ket{e}_k \bra{x} +\mathrm{h.c.},
\end{align}
and we can construct a multi-atom Hamiltonian using an appropriate basis and including interaction terms as follows:
\begin{align}
H_{\mathrm{T}}=\sum^{N_R}_{k=1} I^{2(k-1)}\otimes H_k \otimes I^{2(N_R-k)}+\sum_{i\neq j} H^{\mathrm{Int}}_{ij},
\end{align}
where $I^{2k}$ is the identity matrix of dimension $2k\times 2k $, and $H^{\mathrm{Int}}_{ij}$ describes the Rydberg-Rydberg interaction via
\begin{gather}
H^{\mathrm{Int}}_{ij}=V_{ij}\sigma_{x,x}^i \sigma_{x,x}^j.
\end{gather}

The master equation for the system is $\dot{\rho} = i\hbar\left[H,\rho\right] + \Gamma$ where $\Gamma$ includes radiative damping; for decays from the excited 5P state $\ket{e}$, it takes the form:
\begin{align}
\Gamma = \sum^{N_R+1}_{k=1}\sum_{i=0,1}\frac{1}{2}\gamma_e \left(2\sigma_{i,e}^{k}\rho\sigma_{e,i}^{k}-\sigma_{e,e}^{k}\rho-\rho\sigma_{e,e}^{k}\right),
\end{align}
where $\gamma_e = 6.065 \;\mathrm{MHz}$ is the radiative decay rate for the $5P_{3/2}$ state of $^{87}\mathrm{Rb}$ and $\sigma_{i,j}^k$ is the projector for atom $k$ for states $i$ and $j$. Damping from the excited Rydberg states is characterized by a much-longer lifetime, typically of ms scale, and we may safely neglect the effect for $\mu s$ gate operations. However we do note that including the decays from Rydberg states is more delicate: the atoms are effectively lost from the system, so we would need to trace over any the lost atom(s) in obtaining our final result. Finally, it is possible to include the finite linewidth of the laser beams in our analysis. This is accomplished by the addition of dephasing terms in the master equation, of the form:
\begin{align}
\Gamma = \sum^{N_R+1}_{k=1}\sum_{i=1,e,x}\frac{1}{2}\gamma_{dph} \left(2\sigma_{i,i}^{k}\rho\sigma_{i,i}^{k}-\sigma_{i,i}^{k}\rho-\rho\sigma_{i,i}^{k}\right),
\end{align}
with $\gamma_{dph}$ being the appropriate laser linewidth, and it should be noted that the sum runs over the atoms and the states $\ket{1}$, $\ket{e}$ and $\ket{4}$.

\section{Strongest-pairs approximation}\label{app:spa}
We adopt a simplified treatment of the Rydberg-Rydberg interactions, so that we only consider the strongest interactions between two atoms excited to the $nSn^{\prime}S$ state, and separated by a distance $r$. For the control-ensemble interactions, $n^{\prime}=n-1$, whereas the ensemble atoms have $n^{\prime}=n$. The interaction involves a coupling to a state $n^{\prime}Pn^{\prime\prime}P$ state whose energy is close to that of the $nSn^{\prime}S$ state, differing by an amount $\Delta_{\mathrm{def}}$. For the control-ensemble interactions, $n^{\prime\prime}=n-1$ and for the ensemble-ensemble interactions, $n^{\prime\prime}=n-2$. The energy shift of the doubly excited $nSn^{\prime}S$ state is given by
\begin{gather} V_{C,k}=\frac{1}{2}\left(\Delta_{\mathrm{def}}-\mathrm{sgn}(\Delta_{\mathrm{def}})\sqrt{\Delta^2_{\mathrm{For}} + 4C^2_{\mathrm{dd}}/r^6_k}\right),
\end{gather}
where
\begin{align}\label{eq:cdd}
C_{\mathrm{dd}} = \left(\frac{e}{4\pi\epsilon_0}\right)^2
& \bra{n S n^{\prime}P}r\ket{n S n^{\prime}P} \\\nonumber
& \quad \times \bra{n^{\prime}S n^{\prime\prime}P}r\ket{n^{\prime}S n^{\prime\prime}P}.
\end{align}

\bibliography{1_Qubit_Metrology}

\begin{thebibliography}{21}%
\makeatletter
\providecommand \@ifxundefined [1]{%
 \@ifx{#1\undefined}
}%
\providecommand \@ifnum [1]{%
 \ifnum #1\expandafter \@firstoftwo
 \else \expandafter \@secondoftwo
 \fi
}%
\providecommand \@ifx [1]{%
 \ifx #1\expandafter \@firstoftwo
 \else \expandafter \@secondoftwo
 \fi
}%
\providecommand \natexlab [1]{#1}%
\providecommand \enquote  [1]{``#1''}%
\providecommand \bibnamefont  [1]{#1}%
\providecommand \bibfnamefont [1]{#1}%
\providecommand \citenamefont [1]{#1}%
\providecommand \href@noop [0]{\@secondoftwo}%
\providecommand \href [0]{\begingroup \@sanitize@url \@href}%
\providecommand \@href[1]{\@@startlink{#1}\@@href}%
\providecommand \@@href[1]{\endgroup#1\@@endlink}%
\providecommand \@sanitize@url [0]{\catcode `\\12\catcode `\$12\catcode
  `\&12\catcode `\#12\catcode `\^12\catcode `\_12\catcode `\%12\relax}%
\providecommand \@@startlink[1]{}%
\providecommand \@@endlink[0]{}%
\providecommand \url  [0]{\begingroup\@sanitize@url \@url }%
\providecommand \@url [1]{\endgroup\@href {#1}{\urlprefix }}%
\providecommand \urlprefix  [0]{URL }%
\providecommand \Eprint [0]{\href }%
\providecommand \doibase [0]{http://dx.doi.org/}%
\providecommand \selectlanguage [0]{\@gobble}%
\providecommand \bibinfo  [0]{\@secondoftwo}%
\providecommand \bibfield  [0]{\@secondoftwo}%
\providecommand \translation [1]{[#1]}%
\providecommand \BibitemOpen [0]{}%
\providecommand \bibitemStop [0]{}%
\providecommand \bibitemNoStop [0]{.\EOS\space}%
\providecommand \EOS [0]{\spacefactor3000\relax}%
\providecommand \BibitemShut  [1]{\csname bibitem#1\endcsname}%
\let\auto@bib@innerbib\@empty
\bibitem [{\citenamefont {Giovannetti}\ \emph {et~al.}(2004)\citenamefont
  {Giovannetti}, \citenamefont {Lloyd},\ and\ \citenamefont
  {Maccone}}]{Giovannetti04}%
  \BibitemOpen
  \bibfield  {author} {\bibinfo {author} {\bibfnamefont {V.}~\bibnamefont
  {Giovannetti}}, \bibinfo {author} {\bibfnamefont {S.}~\bibnamefont {Lloyd}},
  \ and\ \bibinfo {author} {\bibfnamefont {L.}~\bibnamefont {Maccone}},\
  }\href@noop {} {\bibfield  {journal} {\bibinfo  {journal} {Science}\ }\textbf
  {\bibinfo {volume} {306}},\ \bibinfo {pages} {1330} (\bibinfo {year}
  {2004})}\BibitemShut {NoStop}%
\bibitem [{\citenamefont {Giovannetti}\ \emph {et~al.}(2011)\citenamefont
  {Giovannetti}, \citenamefont {Lloyd},\ and\ \citenamefont {Maccone}}]{GLM11}%
  \BibitemOpen
  \bibfield  {author} {\bibinfo {author} {\bibfnamefont {V.}~\bibnamefont
  {Giovannetti}}, \bibinfo {author} {\bibfnamefont {S.}~\bibnamefont {Lloyd}},
  \ and\ \bibinfo {author} {\bibfnamefont {L.}~\bibnamefont {Maccone}},\
  }\href@noop {} {\bibfield  {journal} {\bibinfo  {journal} {Nature Photon.}\
  }\textbf {\bibinfo {volume} {5}},\ \bibinfo {pages} {222} (\bibinfo {year}
  {2011})}\BibitemShut {NoStop}%
\bibitem [{\citenamefont {Dowling}(2008)}]{DowlingReview}%
  \BibitemOpen
  \bibfield  {author} {\bibinfo {author} {\bibfnamefont {J.~P.}\ \bibnamefont
  {Dowling}},\ }\href@noop {} {\bibfield  {journal} {\bibinfo  {journal}
  {Contemporary Phys.}\ }\textbf {\bibinfo {volume} {49}},\ \bibinfo {pages}
  {125} (\bibinfo {year} {2008})}\BibitemShut {NoStop}%
\bibitem [{\citenamefont {Caves}\ and\ \citenamefont
  {Shaji}(2010)}]{CavesShaji10}%
  \BibitemOpen
  \bibfield  {author} {\bibinfo {author} {\bibfnamefont {C.~M.}\ \bibnamefont
  {Caves}}\ and\ \bibinfo {author} {\bibfnamefont {A.}~\bibnamefont {Shaji}},\
  }\href@noop {} {\bibfield  {journal} {\bibinfo  {journal} {Opt. Commun.}\
  }\textbf {\bibinfo {volume} {283}},\ \bibinfo {pages} {695} (\bibinfo {year}
  {2010})}\BibitemShut {NoStop}%
\bibitem [{\citenamefont {Afek}\ \emph {et~al.}(2010)\citenamefont {Afek},
  \citenamefont {Ambar},\ and\ \citenamefont {Silberberg}}]{Afek2010}%
  \BibitemOpen
  \bibfield  {author} {\bibinfo {author} {\bibfnamefont {I.}~\bibnamefont
  {Afek}}, \bibinfo {author} {\bibfnamefont {O.}~\bibnamefont {Ambar}}, \ and\
  \bibinfo {author} {\bibfnamefont {Y.}~\bibnamefont {Silberberg}},\
  }\href@noop {} {\bibfield  {journal} {\bibinfo  {journal} {Science}\ }\textbf
  {\bibinfo {volume} {328}},\ \bibinfo {pages} {879} (\bibinfo {year}
  {2010})}\BibitemShut {NoStop}%
\bibitem [{\citenamefont {Demkowicz-Dobrza{\'n}ski}\ \emph
  {et~al.}(2012)\citenamefont {Demkowicz-Dobrza{\'n}ski}, \citenamefont
  {Ko{\l}ody{\'n}ski},\ and\ \citenamefont
  {Gu{\c{t}}{\u{a}}}}]{RafalNatComms12}%
  \BibitemOpen
  \bibfield  {author} {\bibinfo {author} {\bibfnamefont {R.}~\bibnamefont
  {Demkowicz-Dobrza{\'n}ski}}, \bibinfo {author} {\bibfnamefont
  {J.}~\bibnamefont {Ko{\l}ody{\'n}ski}}, \ and\ \bibinfo {author}
  {\bibfnamefont {M.}~\bibnamefont {Gu{\c{t}}{\u{a}}}},\ }\href@noop {}
  {\bibfield  {journal} {\bibinfo  {journal} {Nature Commun.}\ }\textbf
  {\bibinfo {volume} {3:1063}} (\bibinfo {year} {2012})}\BibitemShut {NoStop}%
\bibitem [{\citenamefont {Jones}\ \emph {et~al.}(2009)\citenamefont {Jones},
  \citenamefont {Karlen}, \citenamefont {Fitzsimons}, \citenamefont {Ardavan},
  \citenamefont {Benjamin}, \citenamefont {Briggs},\ and\ \citenamefont
  {Morton}}]{Jones2009}%
  \BibitemOpen
  \bibfield  {author} {\bibinfo {author} {\bibfnamefont {J.~A.}\ \bibnamefont
  {Jones}}, \bibinfo {author} {\bibfnamefont {S.~D.}\ \bibnamefont {Karlen}},
  \bibinfo {author} {\bibfnamefont {J.}~\bibnamefont {Fitzsimons}}, \bibinfo
  {author} {\bibfnamefont {A.}~\bibnamefont {Ardavan}}, \bibinfo {author}
  {\bibfnamefont {S.~C.}\ \bibnamefont {Benjamin}}, \bibinfo {author}
  {\bibfnamefont {G.~A.~D.}\ \bibnamefont {Briggs}}, \ and\ \bibinfo {author}
  {\bibfnamefont {J.~J.~L.}\ \bibnamefont {Morton}},\ }\href@noop {} {\bibfield
   {journal} {\bibinfo  {journal} {Science}\ }\textbf {\bibinfo {volume}
  {324}},\ \bibinfo {pages} {1166} (\bibinfo {year} {2009})}\BibitemShut
  {NoStop}%
\bibitem [{\citenamefont {Modi}\ \emph {et~al.}(2011)\citenamefont {Modi},
  \citenamefont {Cable}, \citenamefont {Williamson},\ and\ \citenamefont
  {Vedral}}]{arXiv:1003.1174}%
  \BibitemOpen
  \bibfield  {author} {\bibinfo {author} {\bibfnamefont {K.}~\bibnamefont
  {Modi}}, \bibinfo {author} {\bibfnamefont {H.}~\bibnamefont {Cable}},
  \bibinfo {author} {\bibfnamefont {M.}~\bibnamefont {Williamson}}, \ and\
  \bibinfo {author} {\bibfnamefont {V.}~\bibnamefont {Vedral}},\ }\href@noop {}
  {\bibfield  {journal} {\bibinfo  {journal} {Phys. Rev. X}\ }\textbf {\bibinfo
  {volume} {1}},\ \bibinfo {pages} {021022} (\bibinfo {year}
  {2011})}\BibitemShut {NoStop}%
\bibitem [{\citenamefont {Cable}\ \emph {et~al.}(2015)\citenamefont {Cable},
  \citenamefont {Gu},\ and\ \citenamefont {Modi}}]{hugo}%
  \BibitemOpen
  \bibfield  {author} {\bibinfo {author} {\bibfnamefont {H.}~\bibnamefont
  {Cable}}, \bibinfo {author} {\bibfnamefont {M.}~\bibnamefont {Gu}}, \ and\
  \bibinfo {author} {\bibfnamefont {K.}~\bibnamefont {Modi}},\ }\href@noop {}
  {\bibfield  {journal} {\bibinfo  {journal} {arXiv:1504.02460}\ } (\bibinfo
  {year} {2015})}\BibitemShut {NoStop}%
\bibitem [{\citenamefont {Modi}\ \emph {et~al.}(2012)\citenamefont {Modi},
  \citenamefont {Brodutch}, \citenamefont {Cable}, \citenamefont {Paterek},\
  and\ \citenamefont {Vedral}}]{Modi12}%
  \BibitemOpen
  \bibfield  {author} {\bibinfo {author} {\bibfnamefont {K.}~\bibnamefont
  {Modi}}, \bibinfo {author} {\bibfnamefont {A.}~\bibnamefont {Brodutch}},
  \bibinfo {author} {\bibfnamefont {H.}~\bibnamefont {Cable}}, \bibinfo
  {author} {\bibfnamefont {T.}~\bibnamefont {Paterek}}, \ and\ \bibinfo
  {author} {\bibfnamefont {V.}~\bibnamefont {Vedral}},\ }\href@noop {}
  {\bibfield  {journal} {\bibinfo  {journal} {Rev. Mod. Phys.}\ }\textbf
  {\bibinfo {volume} {84}},\ \bibinfo {pages} {1655} (\bibinfo {year}
  {2012})}\BibitemShut {NoStop}%
\bibitem [{\citenamefont {Knill}\ and\ \citenamefont
  {Laflamme}(1998)}]{KnillLaflamme98}%
  \BibitemOpen
  \bibfield  {author} {\bibinfo {author} {\bibfnamefont {E.}~\bibnamefont
  {Knill}}\ and\ \bibinfo {author} {\bibfnamefont {R.}~\bibnamefont
  {Laflamme}},\ }\href@noop {} {\bibfield  {journal} {\bibinfo  {journal}
  {Phys. Rev. Lett.}\ }\textbf {\bibinfo {volume} {81}},\ \bibinfo {pages}
  {5672} (\bibinfo {year} {1998})}\BibitemShut {NoStop}%
\bibitem [{\citenamefont {Mansell}\ and\ \citenamefont
  {Bergamini}(2014)}]{mansell}%
  \BibitemOpen
  \bibfield  {author} {\bibinfo {author} {\bibfnamefont {C.}~\bibnamefont
  {Mansell}}\ and\ \bibinfo {author} {\bibfnamefont {S.}~\bibnamefont
  {Bergamini}},\ }\href@noop {} {\bibfield  {journal} {\bibinfo  {journal}
  {New. J. Phys.}\ }\textbf {\bibinfo {volume} {16}},\ \bibinfo {pages}
  {053045} (\bibinfo {year} {2014})}\BibitemShut {NoStop}%
\bibitem [{\citenamefont {M\"uller}\ \emph {et~al.}(2009)\citenamefont
  {M\"uller}, \citenamefont {Lesanovsky}, \citenamefont {Weimer}, \citenamefont
  {B\"uchler},\ and\ \citenamefont {Zoller}}]{Muller2009}%
  \BibitemOpen
  \bibfield  {author} {\bibinfo {author} {\bibfnamefont {M.}~\bibnamefont
  {M\"uller}}, \bibinfo {author} {\bibfnamefont {I.}~\bibnamefont
  {Lesanovsky}}, \bibinfo {author} {\bibfnamefont {H.}~\bibnamefont {Weimer}},
  \bibinfo {author} {\bibfnamefont {H.~P.}\ \bibnamefont {B\"uchler}}, \ and\
  \bibinfo {author} {\bibfnamefont {P.}~\bibnamefont {Zoller}},\ }\href@noop {}
  {\bibfield  {journal} {\bibinfo  {journal} {Phys. Rev. Lett.}\ }\textbf
  {\bibinfo {volume} {102}},\ \bibinfo {pages} {170502} (\bibinfo {year}
  {2009})}\BibitemShut {NoStop}%
\bibitem [{\citenamefont {Braunstein}\ and\ \citenamefont
  {Caves}(1994)}]{CavesBraunstein94}%
  \BibitemOpen
  \bibfield  {author} {\bibinfo {author} {\bibfnamefont {S.~L.}\ \bibnamefont
  {Braunstein}}\ and\ \bibinfo {author} {\bibfnamefont {C.~M.}\ \bibnamefont
  {Caves}},\ }\href@noop {} {\bibfield  {journal} {\bibinfo  {journal} {Phys.
  Rev. Lett.}\ }\textbf {\bibinfo {volume} {72}},\ \bibinfo {pages} {3439}
  (\bibinfo {year} {1994})}\BibitemShut {NoStop}%
\bibitem [{\citenamefont {Bergamini}\ \emph {et~al.}(2004)\citenamefont
  {Bergamini}, \citenamefont {Darqui\'e}, \citenamefont {Jones}, \citenamefont
  {Jacubowiez}, \citenamefont {Browaeys},\ and\ \citenamefont
  {Grangier}}]{bergamini}%
  \BibitemOpen
  \bibfield  {author} {\bibinfo {author} {\bibfnamefont {S.}~\bibnamefont
  {Bergamini}}, \bibinfo {author} {\bibfnamefont {B.}~\bibnamefont
  {Darqui\'e}}, \bibinfo {author} {\bibfnamefont {M.}~\bibnamefont {Jones}},
  \bibinfo {author} {\bibfnamefont {L.}~\bibnamefont {Jacubowiez}}, \bibinfo
  {author} {\bibfnamefont {A.}~\bibnamefont {Browaeys}}, \ and\ \bibinfo
  {author} {\bibfnamefont {P.}~\bibnamefont {Grangier}},\ }\href@noop {}
  {\bibfield  {journal} {\bibinfo  {journal} {J. Opt. Soc. Am. B}\ }\textbf
  {\bibinfo {volume} {21}},\ \bibinfo {pages} {1889} (\bibinfo {year}
  {2004})}\BibitemShut {NoStop}%
\bibitem [{\citenamefont {Reetz-Lamour}\ \emph {et~al.}(2008)\citenamefont
  {Reetz-Lamour}, \citenamefont {Deiglmayr}, \citenamefont {Amthor},\ and\
  \citenamefont {Weidem\"uller}}]{ReetzLamour08}%
  \BibitemOpen
  \bibfield  {author} {\bibinfo {author} {\bibfnamefont {M.}~\bibnamefont
  {Reetz-Lamour}}, \bibinfo {author} {\bibfnamefont {J.}~\bibnamefont
  {Deiglmayr}}, \bibinfo {author} {\bibfnamefont {T.}~\bibnamefont {Amthor}}, \
  and\ \bibinfo {author} {\bibfnamefont {M.}~\bibnamefont {Weidem\"uller}},\
  }\href@noop {} {\bibfield  {journal} {\bibinfo  {journal} {New J. Phys.}\
  }\textbf {\bibinfo {volume} {10}},\ \bibinfo {pages} {045026} (\bibinfo
  {year} {2008})}\BibitemShut {NoStop}%
\bibitem [{\citenamefont {Frese}\ \emph {et~al.}(2000)\citenamefont {Frese},
  \citenamefont {Ueberholz}, \citenamefont {Kuhr}, \citenamefont {Alt},
  \citenamefont {Schrader}, \citenamefont {Gomer},\ and\ \citenamefont
  {Meschede}}]{Frese2000}%
  \BibitemOpen
  \bibfield  {author} {\bibinfo {author} {\bibfnamefont {D.}~\bibnamefont
  {Frese}}, \bibinfo {author} {\bibfnamefont {B.}~\bibnamefont {Ueberholz}},
  \bibinfo {author} {\bibfnamefont {S.}~\bibnamefont {Kuhr}}, \bibinfo {author}
  {\bibfnamefont {W.}~\bibnamefont {Alt}}, \bibinfo {author} {\bibfnamefont
  {D.}~\bibnamefont {Schrader}}, \bibinfo {author} {\bibfnamefont
  {V.}~\bibnamefont {Gomer}}, \ and\ \bibinfo {author} {\bibfnamefont
  {D.}~\bibnamefont {Meschede}},\ }\href@noop {} {\bibfield  {journal}
  {\bibinfo  {journal} {Phys. Rev. Lett.}\ }\textbf {\bibinfo {volume} {85}},\
  \bibinfo {pages} {3777} (\bibinfo {year} {2000})}\BibitemShut {NoStop}%
\bibitem [{\citenamefont {Saffman}\ \emph {et~al.}(2010)\citenamefont
  {Saffman}, \citenamefont {T.Walker},\ and\ \citenamefont
  {M\"olmer}}]{saff10}%
  \BibitemOpen
  \bibfield  {author} {\bibinfo {author} {\bibfnamefont {M.}~\bibnamefont
  {Saffman}}, \bibinfo {author} {\bibnamefont {T.Walker}}, \ and\ \bibinfo
  {author} {\bibfnamefont {K.}~\bibnamefont {M\"olmer}},\ }\href@noop {}
  {\bibfield  {journal} {\bibinfo  {journal} {Rev. Mod. Phys.}\ }\textbf
  {\bibinfo {volume} {82}},\ \bibinfo {pages} {2313} (\bibinfo {year}
  {2010})}\BibitemShut {NoStop}%
\bibitem [{\citenamefont {Mandel}\ \emph {et~al.}(2003)\citenamefont {Mandel},
  \citenamefont {Greiner}, \citenamefont {Widera}, \citenamefont {Rom},
  \citenamefont {H\"ansch},\ and\ \citenamefont {Bloch}}]{Bloch}%
  \BibitemOpen
  \bibfield  {author} {\bibinfo {author} {\bibfnamefont {O.}~\bibnamefont
  {Mandel}}, \bibinfo {author} {\bibfnamefont {M.}~\bibnamefont {Greiner}},
  \bibinfo {author} {\bibfnamefont {A.}~\bibnamefont {Widera}}, \bibinfo
  {author} {\bibfnamefont {T.}~\bibnamefont {Rom}}, \bibinfo {author}
  {\bibfnamefont {T.~W.}\ \bibnamefont {H\"ansch}}, \ and\ \bibinfo {author}
  {\bibfnamefont {I.}~\bibnamefont {Bloch}},\ }\href@noop {} {\bibfield
  {journal} {\bibinfo  {journal} {Phys. Rev. Lett.}\ }\textbf {\bibinfo
  {volume} {91}},\ \bibinfo {pages} {010407} (\bibinfo {year}
  {2003})}\BibitemShut {NoStop}%
\bibitem [{\citenamefont {Schrader}\ \emph {et~al.}(2001)\citenamefont
  {Schrader}, \citenamefont {Kuhr}, \citenamefont {Alt}, \citenamefont
  {M\"{u}ller}, \citenamefont {Gomer},\ and\ \citenamefont
  {Meschede}}]{Meschede}%
  \BibitemOpen
  \bibfield  {author} {\bibinfo {author} {\bibfnamefont {D.}~\bibnamefont
  {Schrader}}, \bibinfo {author} {\bibfnamefont {S.}~\bibnamefont {Kuhr}},
  \bibinfo {author} {\bibfnamefont {W.}~\bibnamefont {Alt}}, \bibinfo {author}
  {\bibfnamefont {M.}~\bibnamefont {M\"{u}ller}}, \bibinfo {author}
  {\bibfnamefont {V.}~\bibnamefont {Gomer}}, \ and\ \bibinfo {author}
  {\bibfnamefont {D.}~\bibnamefont {Meschede}},\ }\href@noop {} {\bibfield
  {journal} {\bibinfo  {journal} {Appl. Phys. B}\ }\textbf {\bibinfo {volume}
  {73}},\ \bibinfo {pages} {819} (\bibinfo {year} {2001})}\BibitemShut
  {NoStop}%
\bibitem [{\citenamefont {Beugnon}\ \emph {et~al.}(2007)\citenamefont
  {Beugnon}, \citenamefont {Tuchendler}, \citenamefont {Marion}, \citenamefont
  {Ga{\"{e}}tan}, \citenamefont {Miroshnychenko}, \citenamefont {Sortais},
  \citenamefont {Lance}, \citenamefont {Jones}, \citenamefont {Messin},
  \citenamefont {Browaeys},\ and\ \citenamefont {Grangier}}]{Browaeys}%
  \BibitemOpen
  \bibfield  {author} {\bibinfo {author} {\bibfnamefont {J.}~\bibnamefont
  {Beugnon}}, \bibinfo {author} {\bibfnamefont {C.}~\bibnamefont {Tuchendler}},
  \bibinfo {author} {\bibfnamefont {H.}~\bibnamefont {Marion}}, \bibinfo
  {author} {\bibfnamefont {A.}~\bibnamefont {Ga{\"{e}}tan}}, \bibinfo {author}
  {\bibfnamefont {Y.}~\bibnamefont {Miroshnychenko}}, \bibinfo {author}
  {\bibfnamefont {Y.~R.~P.}\ \bibnamefont {Sortais}}, \bibinfo {author}
  {\bibfnamefont {A.~M.}\ \bibnamefont {Lance}}, \bibinfo {author}
  {\bibfnamefont {M.~P.~A.}\ \bibnamefont {Jones}}, \bibinfo {author}
  {\bibfnamefont {G.}~\bibnamefont {Messin}}, \bibinfo {author} {\bibfnamefont
  {A.}~\bibnamefont {Browaeys}}, \ and\ \bibinfo {author} {\bibfnamefont
  {P.}~\bibnamefont {Grangier}},\ }\href@noop {} {\bibfield  {journal}
  {\bibinfo  {journal} {Nature Phys.}\ }\textbf {\bibinfo {volume} {3}},\
  \bibinfo {pages} {696} (\bibinfo {year} {2007})}\BibitemShut {NoStop}%
\end{thebibliography}%

\end{document}